\documentclass[preprint]{aastex631}
\accepted{for publication in \apj}

\shorttitle{The Radius Cliff is a Waterfall}
\shortauthors{Chakrabarty, Mulders, Aguichine, \& Batalha}
\usepackage{natbib}
\usepackage{calrsfs}
\usepackage{graphicx}
\usepackage{amsmath}

\begin{document}

\newcommand{\re}{R_\oplus}
\newcommand{\rem}{$R_\oplus$}
\newcommand{\me}{M_\oplus}
\newcommand{\mem}{$M_\oplus$}
\newcommand{\de}{\rho_\oplus}
\newcommand{\dem}{$\rho_\oplus$}
\newcommand{\rs}{R_\odot}
\newcommand{\rsm}{$R_\odot$}
\newcommand{\ms}{M_\odot}
\newcommand{\msm}{$M_\odot$}
\newcommand{\ls}{L_\odot}
\newcommand{\lsm}{$L_\odot$}
\newcommand{\unif}{\mathcal{U}}
\newcommand{\norm}{\mathcal{N}}
\newcommand{\bpoly}{\mathcal{B}}
\newcommand{\f}{\mathcal{F}}
\newcommand{\ho}{H$_{\rm 2}$O}
\newcommand{\simm}{$\sim$}
\newcommand{\rf}{r_f}
\newcommand{\rfm}{$r_f$}
\newcommand{\rw}[1]{#1_r}
\newcommand{\rwm}[1]{$#1_r$}
\newcommand{\ww}[1]{#1_w}
\newcommand{\wwm}[1]{$#1_w$}
\newcommand{\ps}[1]{#1_*}
\newcommand{\psm}[1]{$#1_*$}
\newcommand{\p}{$P$}
\newcommand{\pat}[1]{P_{#1}}
\newcommand{\patm}[1]{$P_{#1}$}
\newcommand{\rp}{R_P}
\newcommand{\rpm}{$R_P$}
\newcommand{\rpat}[1]{R_{P, #1}}
\newcommand{\rpatm}[1]{$R_{P, #1}$}
\newcommand{\wmf}{\rm WMF}
\newcommand{\lij}{\lambda_{ij}}
\newcommand{\lijm}{$\lambda_{ij}$}
\newcommand{\Lij}{\Lambda_{ij}}
\newcommand{\Lijm}{$\Lambda_{ij}$}
\newcommand{\nij}{n_{ij}}
\newcommand{\nijm}{$n_{ij}$}
\newcommand{\cks}{\pi_{cks}}
\newcommand{\cksm}{$\pi_{cks}$}
\newcommand{\lt}{$<$}
\newcommand{\lte}{$\leq$}
\newcommand{\gt}{$>$}
\newcommand{\gte}{$\geq$}
\newcommand{\abin}{\textit{ab initio}}

\title{The Radius Cliff is a Waterfall: Explaining Sub-Neptune Exoplanets with Steam Worlds}

\correspondingauthor{Aritra Chakrabarty}
\email{aritra.astrophysics@gmail.com}

\author[0000-0001-6703-0798]{Aritra Chakrabarty}
\altaffiliation{NASA Postdoctoral Program Fellow}
\affil{NASA Ames Research Center, Moffett Field, CA, USA}

\author[0000-0002-1078-9493]{Gijs D. Mulders}
\affil{Instituto de Astrof\'isica, Pontificia Universidad Cat\'olica de Chile, Av. Vicu\~na Mackenna 4860, 7820436 Macul, Santiago, Chile}

\author[0000-0002-8949-5956]{Artyom Aguichine}
\affil{Department of Astronomy and Astrophysics, University of California, Santa Cruz, CA, USA}
\affil{Instituto de Astronom\'ia, Universidad Nacional Aut\'onoma de M\'exico, Apartado Postal 106, CP 22800 Ensenada, Baja California, M\'exico}

\author[0000-0002-7030-9519]{Natalie Batalha}
\affil{Department of Astronomy and Astrophysics, University of California, Santa Cruz, CA, USA}



\begin{abstract}

The demographics of Kepler planets provide a key testbed for models of planet formation and evolution, particularly for explaining the \textit{radius valley} separating super-Earths and sub-Neptunes. A primordial interpretation based on differences in bulk densities—where rocky and water-rich planets form via migration pathways—offers an alternative to atmospheric loss scenarios. Updated interior structure models of water worlds with adiabatic steam atmospheres reproduce the observed valley near $\sim2,R_\oplus$ more accurately. Furthermore, migration models from our \textsc{Genesis} library suggest that these formation pathways can also account for the distinct period distributions of super-Earths and sub-Neptunes, as well as the emergence of the hot Neptune desert. Motivated by this, we develop a Bayesian hierarchical mixture model for close-in Kepler planets ($P<100$ days), combining rocky planets and water worlds without H/He envelopes. The inferred mass distributions of rocky and water-rich planets peak at $\sim2.6 \me$ and $\sim7 \me$, respectively, with the water mass fraction of water worlds peaking at $\sim41\%$. Water worlds provide a good representation of the Kepler sub-Neptune population, with the \textit{radius cliff} emerging as a “waterfall”—a sharp decline in their occurrence. However, our mass–radius analysis shows that water worlds alone cannot explain planets with $R\gtrsim3,R_\oplus$, implying that at least $\sim20\%$ of sub-Neptunes in the sample are enriched in H/He gas.

\end{abstract}


\section{Introduction} \label{sec:intro}

The Kepler mission has revealed a rich and diverse population of small, close-in exoplanets. The distinct patterns observed within this population reflect underlying differences in planetary composition, formation, or evolutionary history, and offer critical constraints for models of planet formation and atmospheric evolution. Among the most striking features are the \textit{radius valley}, a deficit of planets $\sim 1.8~\re$ \citep{fulton17, fulton18, vaneylen19, ho23}, which divides the population into super-Earths and sub-Neptunes, and the \textit{radius cliff} near $4~\re$ that marks a sharp drop in the occurrence of small planets \citep{fulton17, fulton18, dattilo23, dattilo24}. Another closely related feature is the lack of ultra-short-period ($P \lesssim 10$ days) sub-Neptunes and Neptune-sized planets, known as the \textit{hot Neptune desert} \citep{mazeh16}.

Atmospheric loss models \citep{owen17, ginzburg18, gupta19, owen20, rogers21} explain the radius valley as a consequence of evolutionary processes, wherein initially gaseous sub-Neptunes, also called gas dwarfs, lose their H/He envelopes and are stripped down to bare rocky cores. While successful in many respects, these models work well with an \textit{in situ} formation scenario \citep{owen17, rogers21}, which is broadly inconsistent with planet formation theories \citep{ogihara15}. Furthermore, they require additional mechanisms to account for the radius cliff near $\sim 4~\re$, such as spontaneous mass loss of H/He envelopes \citep{ikoma12, ginzburg16, owen16}, sequestration of H$_2$ in magma oceans \citep{kite19}, among others. Alternatively, formation theories propose that the observed dichotomy may arise from intrinsic differences in bulk composition---namely, rocky super-Earths versus water-ice-rich sub-Neptunes--- shaped by accretion and migration processes in the early protoplanetary disk \citep{mordasini09, izidoro17, venturini20, mulders20, izidoro22}. This removes the need for all sub-Neptunes to possess significant H/He-rich envelopes to begin with.

Initial water world models assumed water as a condensed, high-pressure ice layer \citep{zeng19}, resulting in planets less ``puffy" than gas dwarfs. Population models based on such structures often struggle to match observations, typically placing water worlds either at the radius valley \citep{venturini20, burn24} or in the super-Earth regime \citep{chakrabarty24}. In contrast, models invoking hot, adiabatic, supercritical steam layers for close-in water worlds offer a more natural explanation for the radius valley, as steam atmospheres substantially inflate planetary radii \citep{turbet20, aguichine21}. \citet{burn24} showed that their migration-based planet formation model can reproduce the $\lesssim 4~\re$ Kepler population when it includes water worlds with mixed steam and hydrogen atmospheres undergoing photoevaporation. However, in the absence of atmospheric escape, the model underproduces rocky super-Earths and fails to generate a sufficient number of water worlds beyond $\sim 3~\re$ (see Figure~2c of \citet{burn24}).

This raises an important question for formation models: Can a primordial population of rocky planets and water-rich worlds reproduce the observed Kepler demographics without invoking post-formation atmospheric evolution? Moreover, can formation and migration processes inherently reproduce the transition from the hot Neptune desert regime to the warm sub-Neptune-dominated regime across orbital periods \citep{fulton17, fulton18, petigura18, pascucci19, bergsten22}---a feature that atmospheric loss models naturally reproduce, providing compelling evidence in their favor \citep{owen17, rogers21}? Addressing these questions can help guide the development of future formation-based population synthesis models.

In this work, we present a hierarchical population model of rocky planets and water-rich steam worlds devoid of H/He envelopes. Using a Bayesian inference framework, we analyze the period–radius distribution of close-in Kepler planets (orbital periods $< 100$ days) from the California-Kepler Survey \citep[CKS,][]{johnson17, petigura17}, constraining the underlying period, mass, and water mass-fraction distributions that best reproduce the observed features. We further assess our inferred model by comparing it with the observed mass–radius distribution and with mass distributions derived from independent radial velocity (RV) surveys. Our results offer constraints for future formation models aiming to explain the observed exoplanet demographics and highlight how future observations can help narrow down the processes that likely shaped the Kepler population.

\S\ref{sec:meth} outlines our methodology, including the definition of the mass, orbital period, and water mass-fraction distributions for rocky planets and water worlds, and how we infer the model parameters by maximizing the likelihood of detecting the Kepler planets under our framework. \S\ref{sec:res} presents the results of our Bayesian inference. \S\ref{sec:future} discusses how our model can inform and guide future observational efforts. Finally, we summarize the key conclusions in \S\ref{sec:con}.

\section{Method}\label{sec:meth}

We develop a Bayesian hierarchical model of planet populations based on the theory of \abin~dichotomy in the planetary bulk compositions, to represent the small, short-period planets observed by Kepler. Accordingly, we assume that the Kepler super-Earths are primarily rocky, and the sub-Neptunes are composed of water-ice-rich cores. For the water worlds, we adopt an updated internal structure model that extends the previous isothermal condensed water-layer model \citep[e.g.,][]{zeng19} by incorporating supercritical and extended steam atmospheres. These ``steam worlds” \citep{aguichine21, burn24} are physically larger than their condensed counterparts and represent promising candidates for explaining the observed radii of Kepler sub-Neptunes. All planets in our model are assumed to be bare, i.e., devoid of primordial H/He envelopes.

We then use a Bayesian framework to fit our model to the observed period–radius (\p-\rpm) distribution of Kepler planets with radii between 0.9–6 \rem~and period (\p) $<$ 100 days. Through this inference, we aim to constrain the following aspects of the planet population:

\begin{itemize}
    \item The mass distributions of rocky planets (\rwm{M}) and water worlds (\wwm{M}).
    \item The water mass-fraction (WMF) distribution of the water worlds.
    \item The period distributions of the rocky planets (\rwm{P}) and water worlds (\wwm{P}).
    \item The occurrence rate of planets per star (\psm{f}) within the range of $0.9~\re < \rp < 6~\re$ and $P < 100$ days.
\end{itemize}

\subsection{Distributions of Period, Mass, and Water Mass-Fraction in the Population Model} \label{sec:meth/dists}

As a first step in our hierarchical mixture model, we define the distributions of orbital period, mass, and WMF from which we draw planet samples. Each of the rocky and water-rich populations is assigned its own conditional distributions of period, mass, and WMF, as explained in the following subsections.

\subsubsection{Orbital period distributions}\label{sec:meth/dists/pdist}

To define the period distribution of the ensemble population, we adopt the smooth broken power-law form from \citet{rogers21}, which fits the observed Kepler occurrence rates:

\begin{equation} \label{eq:pkep}
\frac{dN}{d\log P} \propto \frac{1}{(\frac{P}{P_0})^{-k_1} + (\frac{P}{P_0})^{-k_2}}
\end{equation}


The overall period distribution is partitioned into composition-based conditional distributions for rocky planets and water worlds, motivated by the observed trend that super-Earths dominate at shorter orbital periods while sub-Neptunes become more common at longer periods \citep{fulton18, petigura18, pascucci19, bergsten22}. Our model assumes that this difference is a primordial feature shaped by formation and migration processes, which is further discussed and justified in \S\ref{sec:res/pdist}. We therefore adopt empirical fits to the bias-corrected Kepler period distributions reported by \citet{bergsten22}. These fits provide a functional form for the fractional occurrence of super-Earths as a function of period, $G(P)$, from which we define the conditional period distributions of rocky planets and water worlds as:
\begin{align} \label{eq:prpw}
    \frac{dN}{d\log P_r} \propto G \frac{dN}{d\log P} \nonumber \\
    \frac{dN}{d\log P_w} \propto (1-G) \frac{dN}{d\log P}
\end{align}

G(P) is defined by \citet{bergsten22} via a hyperbolic tangent transition function:
\begin{equation} \label{eq:t-p}
t(P) = 0.5 - 0.5 \tanh \left(\frac{\log P - \log P_{\mathrm{cen}}}{\log s}\right)
\end{equation}
with
\begin{equation} \label{eq:G-t}
G(P) = t(P) \cdot \chi_1 + (1 - t(P)) \cdot \chi_2
\end{equation}

Here, $P_{cen}$ is the transition center, $s$ controls the smoothness of the transition, and $\chi_1$ and $\chi_2$ are the limiting fractions of super-Earths at short and long periods, respectively. 

Since G(P) is defined as the fractional occurrence of super-Earths within the ensemble, it also allows us to calculate the global fractions of rocky planets (\rfm) and water worlds (1-\rfm), where \rfm~can be calculated as
\begin{equation}\label{eq:rf}
    \rf = \frac{\int{G(P)\frac{dN}{d\log P}d\log P}}{\int{\frac{dN}{d\log P}d\log P}}
\end{equation}

\subsubsection{Mass Distributions}\label{sec:meth/dists/mdist}

Choosing an appropriate mass distribution for the Kepler planets is challenging. While a subset of Kepler planets have mass measurements, and some broader trends in planetary masses have emerged \citep[][etc.]{howard10, cassan12, marcy14, fulton21} from independent radial velocity and microlensing surveys (often probing different planet populations), a detailed and completeness-corrected mass distribution specifically for Kepler super-Earths and sub-Neptunes remains unavailable. Several studies have instead used forward population modeling to infer the mass distribution indirectly. For example, \citet{wu19, rogers21} showed that the observed Kepler period-radius distribution can be reproduced using a log-normal distribution of masses.

However, we find that applying a single log-normal mass distribution across the full planet population leads to poor convergence of our model, primarily due to contamination of the rocky component with implausibly large masses. We therefore adopt separate log-normal mass distributions for rocky planets (\rwm{M}) and water worlds (\wwm{M}), allowing greater flexibility in capturing the underlying physical differences between the two populations:
\begin{align} \label{eq:mrmw}
    \frac{dN}{d\log M_r} &= \norm(\log \mu_{mr}, \sigma_{mr}), \nonumber \\
    \frac{dN}{d\log M_w} &= \norm(\log \mu_{mw}, \sigma_{mw})
\end{align} 

This choice of separate mass distributions not only improves the stability of the model fit but is also supported by observed differences in the mass–radius relationships of super-Earths and sub-Neptunes \citep{wolfgang16, chen17}. The resulting combined mass distribution remains of particular interest and is further discussed in \S\ref{sec:res/mwfdist}.

\subsubsection{WMF Distributions}\label{sec:meth/dists/wfdist}

The WMF is set to zero for rocky planets, consistent with a dry, Earth-like bulk composition. For the water worlds, however, the WMF distribution is not constrained by models or observations. Therefore, the first three models we test correspond to three different WMF distributions for the water worlds:
\begin{itemize}
    \item Model I: A truncated normal distribution, $\mathcal{N}(\mu_{\mathrm{WMF}}, \sigma_{\mathrm{WMF}})$, restricted to [0, 0.5];
    \item Model II: A uniform distribution, $\mathcal{U}(a_{\mathrm{WMF}}, b_{\mathrm{WMF}})$, with $b_{\mathrm{WMF}}<0.5$; 
    \item Model III: A non-parametric model based on a fifth-order Bernstein polynomial expansion, normalized over the range (0, 0.5).
\end{itemize}

We truncate the WMF at 0.5 as a pragmatic, physically motivated upper limit. Condensation chemistry and the elemental budget of a solar-composition nebula suggest that water ice constitutes $\lesssim 50 \%$ of the solid mass \citep{lodders03, marboeuf14}. In addition, processes such as radial transport \citep[pebble drift and sublimation at the snowline;][]{booth17} and post-accretion evolution \citep[e.g., differentiation and volatile loss;][]{genda03} typically reduce the water ultimately incorporated into planets to well below this ceiling. 

Furthermore, to assess whether this truncation on WMF affects the MCMC convergence, we also develop Model IV, which considers a normal distribution of WMF like Model I but untruncated, i.e., $0 <$ WMF $< 1$. We evaluate each of the proposed WMF models within our Bayesian framework, as detailed in the following sections. The rationale behind our final choice of WMF parameterization is discussed in \S\ref{sec:meth/modelcomp}.

\subsection{Properties of the host stars} \label{sec:meth/host}

We adopt Sun-like central stars, in both mass and luminosity, for Models I–IV. To investigate the effect of host star properties, we develop Model V, in which the planets from Model I are placed around Kepler-like host stars. The stellar mass (\psm{M}) and luminosity (\psm{L}) for these stars are directly drawn from the CKS stellar sample via bootstrapping (with replacement).

\subsection{Calculating Size of the Planets} \label{sec:meth/rp}

\begin{figure}
\centering
\includegraphics[scale=0.4]{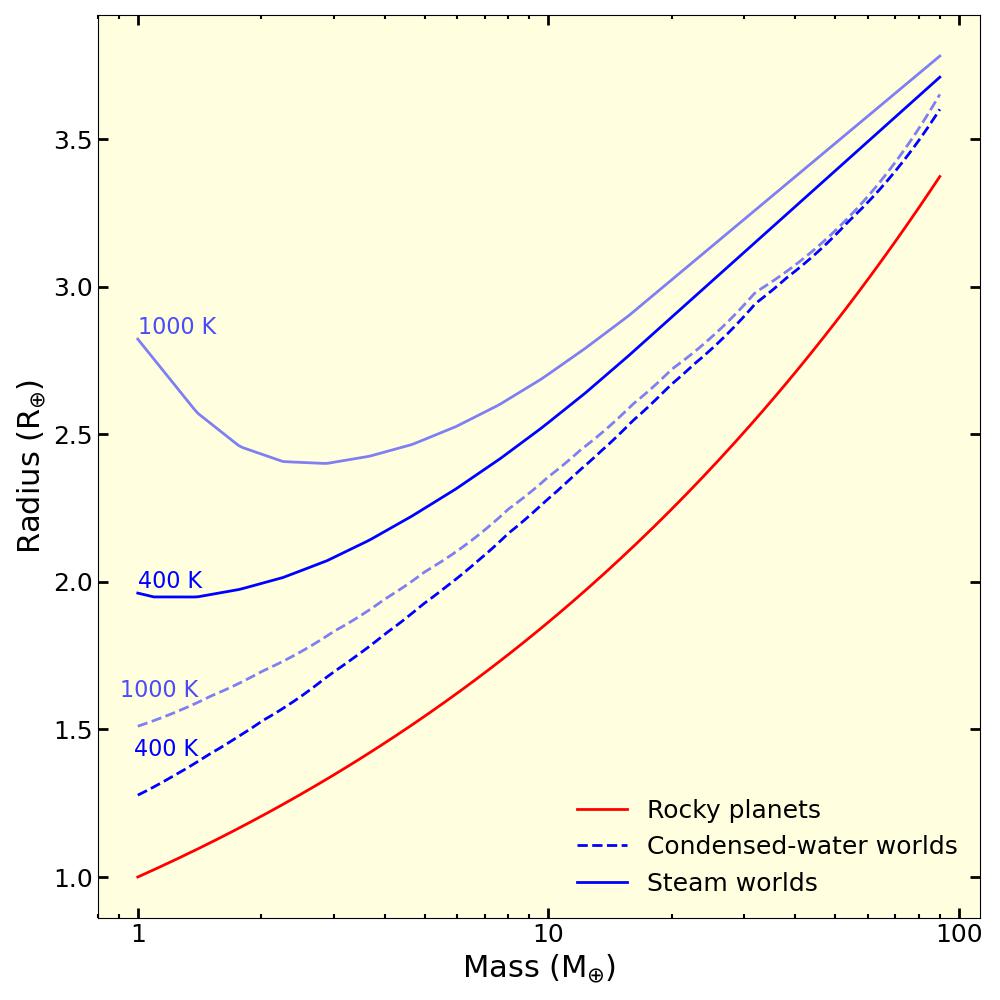}
\caption{Mass–radius models for rocky planets \citep{zeng19} and water worlds with WMF $= 0.5$, having condensed-water layers \citep{zeng19} and steam atmospheres \citep{aguichine21}.
}
\label{fig:mrmodel}
\end{figure}

We assume an Earth-like rocky composition for the rocky planets with 32.5\% iron and 67.5\% silicates by mass. We then use the mass-radius relation from \cite{zeng19}, $m=r^{3.7}$, where both the mass (m) and size (r) are in Earth units. For the water worlds, we assume the same core composition and consider an adiabatic layer of \ho~(steam) on top. To calculate the sizes of these water (steam) worlds, we adopt the equations of state and the static internal structure model of \cite{aguichine21}, in which the entropy of the steam envelope does not evolve. This model provides mass–radius relationships for such planets as a function of core fraction, WMF, and irradiation temperature. We interpolate from their publicly available model grid\footnote{\url{https://archive.lam.fr/GSP/MSEI/IOPmodel/mr_all.dat}} to calculate the radii of our water worlds. Figure~\ref{fig:mrmodel} illustrates the mass–radius models for rocky planets and water worlds with WMF $= 0.5$, where steam atmospheres expand relative to the condensed-water model with increasing equilibrium temperature.

\subsection{Calculation of Likelihood for Bayesian Inference} \label{sec:meth/likelihood}

To calculate the occurrence rates from our synthetic planet population, we first compute the probability density function (PDF) of occurrence as a function of radius and orbital period of the planets using 2D Gaussian Kernel Density Estimation (KDE) technique (implemented via the Python package \textit{scipy.stats.gaussian\_kde}). We calculate the KDE of the population ($p_{kde}$) in the period-radius space on a logarithmic scale over a defined grid of periods and radii. The KDE bandwidth is set to the package’s baseline value, determined using Scott's rule. We additionally explore alternative bandwidths by scaling this baseline up and down, and find that our results are largely insensitive to the choice of bandwidth within a reasonable range, as discussed in Section~\ref{sec:meth/modelcomp}.  We then calculate the occurrence rate per star, $\lambda(P, \rp)$, by multiplying $p_{kde}$ with \psm{f}, the average number of planets per star within the range $0.9~\re < \rp < 6~\re$ and $P < 100$ days.

We then introduce biases associated with the CKS with the help of its completeness map. We follow the approach of \cite{fulton17, fulton18} and partly use their code \textit{cksgaia}\footnote{\url{https://github.com/California-Planet-Search/cksgaia}} for this purpose. Accordingly, we take the product of the transit probability and the detection probability to calculate the completeness factor at the same grid points over which we calculate the KDE. \textit{cksgaia} calculates the transit probability as
\begin{equation} \label{eq:ptran}
    p_{transit} = b_{cut}\frac{R_*}{a}
\end{equation}

$R_*$ is the stellar radius, $a$ is the semi-major axis, and $b_{cut}=0.7$ is the cutoff in the transit parameter \citep{fulton17, fulton18}. The mean probability of detection resulting from the Kepler pipeline efficiency is then calculated as
\begin{equation} \label{eq:pdet}
    p_{detection} = \frac{1}{N_*}\sum_k^{N_*}C(m_k)
\end{equation}

\psm{N}~denotes the number of stars in the CKS sample, and $C(m_i)$ denotes the fraction of injected planets that can be recovered using the Kepler pipeline as a function of its MES, i.e., the injected signal-to-noise ratio. $C(m_i)$ has been calculated using a Gamma cumulative distribution function (see Equations 2 and 3 of \cite{fulton17}). The total completeness function is given by:
\begin{equation} \label{eq:eta}
    \eta = p_{transit} \cdot p_{detection}
\end{equation}

Thus, the average number of detectable planets in a grid cell ($i,j$), with a grid spacing of $\Delta P$ and $\Delta\rp$, is given by
\begin{equation} \label{eq:Lij}
    \Lij = \ps{N} \lij \eta_{ij} \Delta P~\Delta\rp
\end{equation}

We assume that the exoplanets can be detected independently from each other \citep{bryson20, bryson21, rogers21}. Hence, we treat the detection of an exoplanet as a Poisson point process. The likelihood of detecting the ensemble of planets (\cksm) from the CKS sample given the free parameters ($\Theta$) can be expressed as (see \cite{bryson20})
\begin{equation} \label{eq:likelihood}
    p(\cks|\Theta) = \prod_{i,j \in \cks}Poisson(\nij=1, \Lij(\Theta)) \prod_{i,j \notin \cks}Poisson(\nij=0, \Lij(\Theta))
\end{equation}

\nijm~denotes the number of planets detected in the grid cell ($i,j$). $Poisson(n, \Lambda)$ denotes the Poisson probability density function with mean $\Lambda$.

Thus, finding the free parameters consistent with the CKS sample translates to maximizing the posterior probability that the current set of parameters represents the population in the CKS sample, $p(\Theta|\cks)$. Following Bayes' theorem, this can be expressed,  as
\begin{equation} \label{eq:posterior}
    p(\Theta|\cks) \propto p(\cks|\Theta)~p(\Theta) 
\end{equation}

$p(\Theta)$ is a prior on $\Theta$, which is explained in detail in the following subsection. 

\subsection{Maximizing the posterior likelihood using MCMC} \label{sec:meth/mcmc}

We maximize Equation~\ref{eq:posterior} using Markov Chain Monte Carlo (MCMC), implemented via the Python package \textit{emcee}, to obtain the posterior distributions of the free parameters ($\Theta$). The parameters $P_0$, $k_1$, and $k_2$ in Equation~\ref{eq:pkep} only dictate the period distribution of the ensemble population. Similarly, the parameter $P_{\mathrm{cen}}$ in Equation~\ref{eq:t-p} controls only the period at which the transition from super-Earth–dominated to sub-Neptune–dominated populations occurs. These parameters can therefore be constrained directly from the period distribution alone, and including them in the joint fit to the period–radius distribution only slows the convergence of the MCMC across all models without affecting our conclusions. Accordingly, rather than re-inferring them, we adopt values from the literature. We take $P_0$, $k_1$, and $k_2$ from \cite{rogers21}, who demonstrated that these parameters successfully reproduce the Kepler period distribution. We fix $P_{\rm{cen}} = 11.33$ days, corresponding to the median value reported by \citet{bergsten22} for stars in the mass range of 0.56–1.63~\msm (see Table~2 of \cite{bergsten22}).

Thus, the free parameters ($\Theta$) of our model are:
\begin{itemize}
\item Parameters associated with the mass distributions: $\mu_{mr}$, $\sigma_{mr}$, $\mu_{mw}$, and $\sigma_{mw}$
\item Parameters associated with the WMF distributions: i) $\mu_{\wmf}$ and $\sigma_{\wmf}$ for the normal distribution, ii) $a_{\wmf}$ and $b_{\wmf}$ for the uniform distributions, and iii) the Bernstein coefficients, $\beta_{k,\wmf}~(k=1,..,5)$, for the Bernstein polynomial-based distribution.
\item Parameters governing the period distributions: $s$, $\chi_1$, and $\chi_2$
\item Parameter related to the ensemble: \psm{f}
\end{itemize}

We assign uniform priors to all free parameters. The priors for the free parameters associated with the mass distributions are chosen empirically. The WMF distribution parameters are also chosen empirically depending on the distribution type, with the constraint $0<\wmf<0.5$ ($0<\wmf<1$ in the case of Moel IV). The prior for the parameter $s$ is guided by \citet{bergsten22}, while $\chi_1$ and $\chi_2$ are assigned uniform priors over the full range (0, 1). The prior for \psm{f} is informed by a rough estimate obtained by dividing the sum of completeness weights (1/$\eta$) computed for the planets in the CKS sample within our region of interest by \psm{N}, thereby providing a physically motivated constraint. 

\begin{figure}
\centering
\includegraphics[scale=0.3]{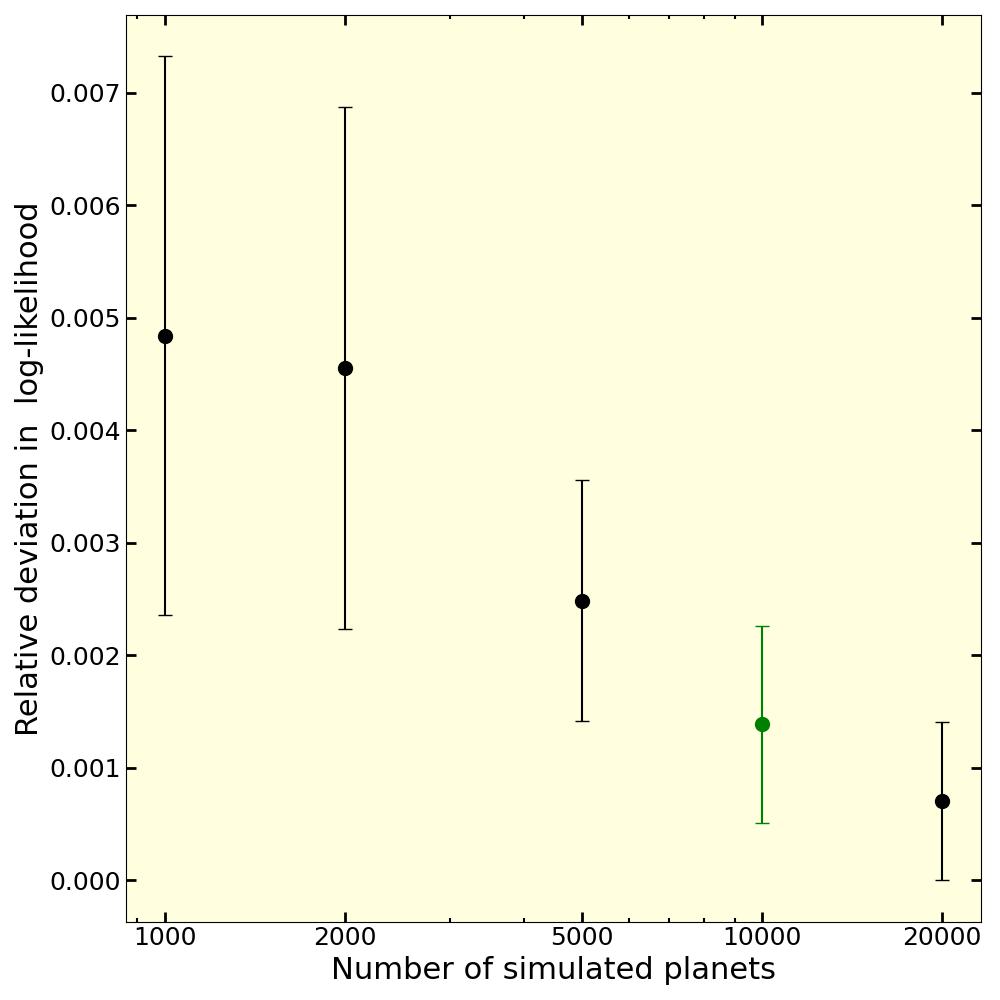}
\caption{Relative deviation of the log-likelihood values for different numbers of simulated planets with respect to the median log-likelihood obtained from 100,000 simulated planets. The green point denotes our chosen number of simulated planets, i.e., 10,000, with which we get $<$0.2\% deviation while maintaining a manageable computational cost.
}
\label{fig:nsim}
\end{figure}

We simulate 10,000 planets in each run to accurately compute the detection likelihood from our model. Figure~\ref{fig:nsim} shows the relative deviation of the log-likelihood ($\log p(\cks|\Theta)$) from the median value obtained from 100,000 simulated planets, plotted as a function of the number of simulated planets. Using 10,000 planets achieves adequate accuracy without excessive computational expense.

\subsection{Comparison of Predictive Performance of the Models}\label{sec:meth/modelcomp}

\begin{figure}
\centering
\includegraphics[scale=0.55]{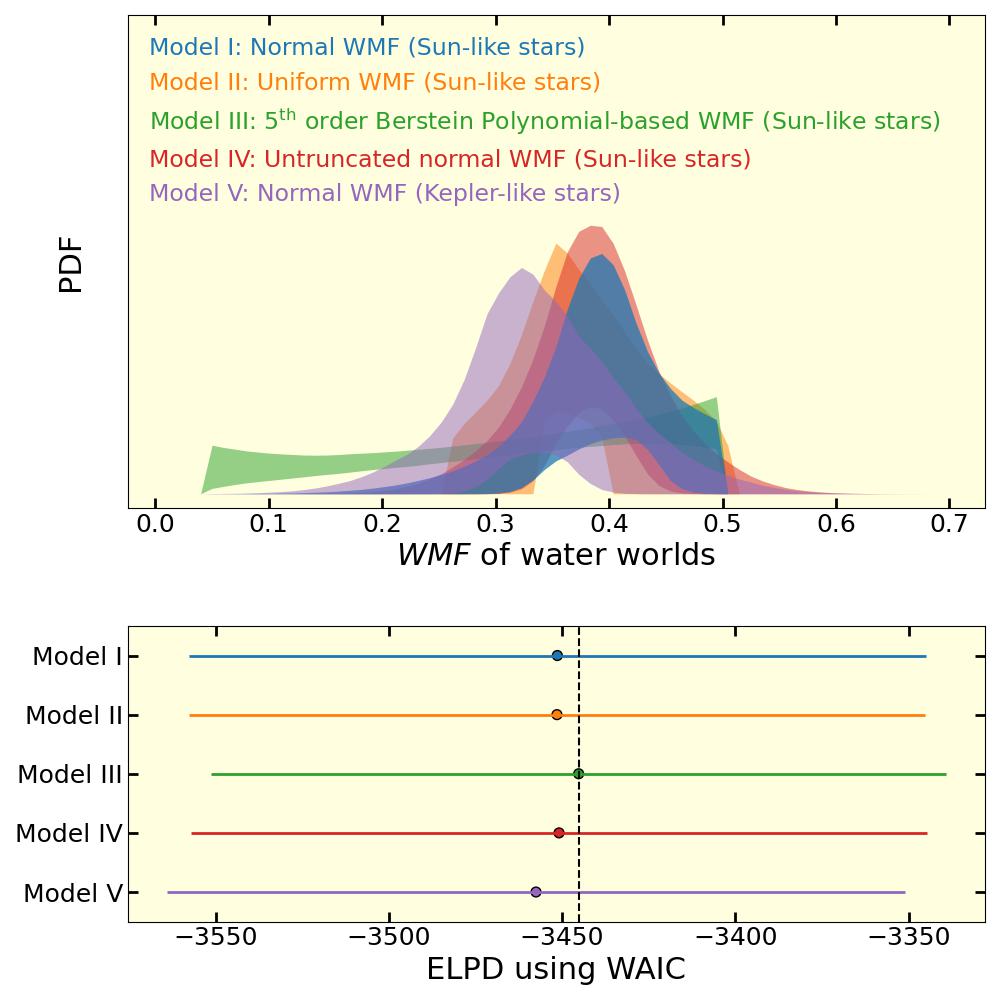}
\caption{Top: Shaded regions indicate the 1-$\sigma$ range of WMF distributions from the different models. Bottom: ELPD values and their associated uncertainties for the different models. The ELPD values show significant overlap, with Model I slightly outperforming the others.
}
\label{fig:arviz}
\end{figure}

We evaluate the performance of models I-V presented in \S\ref{sec:meth/dists} and \S\ref{sec:meth/host} by computing the expected log pointwise predictive density (ELPD) using the widely applicable information criterion (WAIC) method, along with its uncertainty, with the help of the Python package \textit{ArviZ}. ELPD provides a robust estimate of a model’s predictive accuracy by summing the expected log-likelihoods for individual data points. WAIC offers a Bayesian approximation to ELPD by incorporating the full posterior distribution and penalizing model complexity, making it suitable for comparing models in a probabilistic framework \citep{watanabe10, vehtari17}. Figure~\ref{fig:arviz} shows the $1\sigma$ range of the posterior WMF distributions from the four models (top) and a comparison of their ELPD values and uncertainties (bottom). 

The results indicate that the ELPD values for all five models largely overlap. Notably, the $1\sigma$ range of WMF distributions from Model II resembles a normal distribution similar to that from Model I, rather than a uniform distribution, indicating a preferred WMF value inferred from the data. Again, the untruncated normal WMF model in Model IV shows that the peak location and overall WMF distribution closely resemble those of Model I, except for being truncated at the right edge. The ELPD value is also similar, suggesting that the applied truncation on the WMF distribution---motivated by disk chemistry and accretion processes---barely affects the MCMC convergence. Again, adopting Kepler-like central stars in Model V causes the WMF distribution peak at a slightly lower value than Model I with sun-like stars, albeit with 1$\sigma$ overlap with Model I. However, their predictive performances, measured by the ELPD, are statistically indistinguishable.

Furthermore, we find that the predictive performance of Model I is robust to the choice of KDE bandwidth within a reasonable range. We repeat the analysis for Model I by scaling the baseline bandwidth by factors of 0.25, 0.5, and 2, and find that both the ELPDs and the inferred parameters remain very close to their baseline values. For a bandwidth scaling factor of 4, however, the MCMC chains fail to converge, producing a NaN value for the ELPD. Accordingly, we adopt the baseline KDE bandwidth for all results presented in the following section.
\\

\section{Results and Discussion} \label{sec:res}

The posterior mean and uncertainties of the free parameters from MCMC for Models I-V are shown in Table~\ref{tab:params}. The inferred free parameters are highly consistent across models. Although Models I–IV differ by construction in their WMF distribution parameterizations, the inferred WMF distributions are nonetheless broadly similar across models, with the exception of Model III (see Figure~\ref{fig:arviz}). We therefore adopt Model I as our fiducial model and present all subsequent results based on Model I. The full MCMC posterior distribution for Model I is shown in Appendix Figure~\ref{fig:corner}.

\begin{deluxetable}{cccccc}[!h]
\tablecaption{}
\label{tab:params}
\tablewidth{0pt}
\tablehead{
\colhead{Parameters} & \colhead{Model I} & \colhead{Model II} & \colhead{Model III} & \colhead{Model IV} & \colhead{Model V}
}
\startdata
$\mu_{mr}$ & $2.65 \pm 0.14$ & $2.62 \pm 0.14$ & $2.52 \pm 0.15$ & $2.65 \pm 0.14$ & $2.62 \pm 0.15$ \\
$\sigma_{mr}$ & $0.74 \pm 0.06$ & $0.74 \pm 0.06$ & $0.71 \pm 0.06$ & $0.74 \pm 0.06$ & $0.73 \pm 0.07$ \\
$\mu_{mw}$ & $6.99 \pm 2.36$ & $7.76_{-2.78}^{+2.13}$ & $9.41_{-3.8}^{+1.95}$ & $7.37_{-2.64}^{+2.01}$ & $6.2_{-2.07}^{+2.97}$ \\
$\sigma_{mw}$ & $1.82_{-0.16}^{+0.22}$ & $1.8_{-0.16}^{+0.22}$ & $1.71_{-0.16}^{+0.29}$ & $1.81_{-0.17}^{+0.24}$ & $1.83_{-0.17}^{+0.14}$ \\
$\mu_{\wmf}$ & $0.41 \pm 0.05$ & -- & -- & $0.39 \pm 0.03$ & $0.34 \pm 0.05$ \\
$\sigma_{\wmf}$ & $0.06 \pm 0.05$ & -- & -- & $0.04 \pm 0.03$ & $0.04 \pm 0.04$ \\
$a_{\wmf}$ & -- & $0.31 \pm 0.05$ & -- & -- & -- \\
$b_{\wmf}$ & -- & $0.45 \pm 0.06$ & -- & -- & -- \\
$\beta_{0,\wmf}$ & -- & -- & $0.2_{-0.15}^{+0.35}$ & -- & -- \\
$\beta_{1,\wmf}$ & -- & -- & $0.19_{-0.15}^{+0.33}$ & -- & -- \\
$\beta_{2,\wmf}$ & -- & -- & $0.34_{-0.25}^{+0.38}$ & -- & -- \\
$\beta_{3,\wmf}$ & -- & -- & $0.63_{-0.38}^{+0.27}$ & -- & -- \\
$\beta_{4,\wmf}$ & -- & -- & $0.74_{-0.33}^{+0.19}$ & -- & -- \\
\psm{f} & $1.12 \pm 0.05$ & $1.12 \pm 0.04$ & $1.25 \pm 0.05$ & $1.11 \pm 0.04$ & $1.11 \pm 0.05$ \\
$s$ & $3.29_{-0.62}^{+0.7}$ & $3.4_{-0.6}^{+0.73}$ & $3.24_{-0.58}^{+0.5}$ & $3.29_{-0.63}^{+0.75}$ & $3.29_{-0.61}^{+0.95}$ \\
$\chi_1$ & $0.97_{-0.04}^{+0.02}$ & $0.98_{-0.03}^{+0.02}$ & $0.97_{-0.05}^{+0.02}$ & $0.97_{-0.04}^{+0.02}$ & $0.96_{-0.04}^{+0.03}$ \\
$\chi_2$ & $0.22 \pm 0.06$ & $0.21 \pm 0.06$ & $0.20 \pm 0.06$ & $0.21_{-0.04}^{+0.07}$ & $0.22 \pm 0.06$ \\
$\rf$ & $0.47 \pm 0.04$ & $0.47 \pm 0.04$ & $0.46 \pm 0.03$ & $0.47 \pm 0.04$ & $0.47 \pm 0.04$
\enddata
\tablecomments{$\rf$ is derived from the posterior sample of the free parameters for each model using Equation~\ref{eq:rf}}
\end{deluxetable}

\subsection{The period distributions}\label{sec:res/pdist}

\begin{figure}
\centering
\includegraphics[scale=0.5]{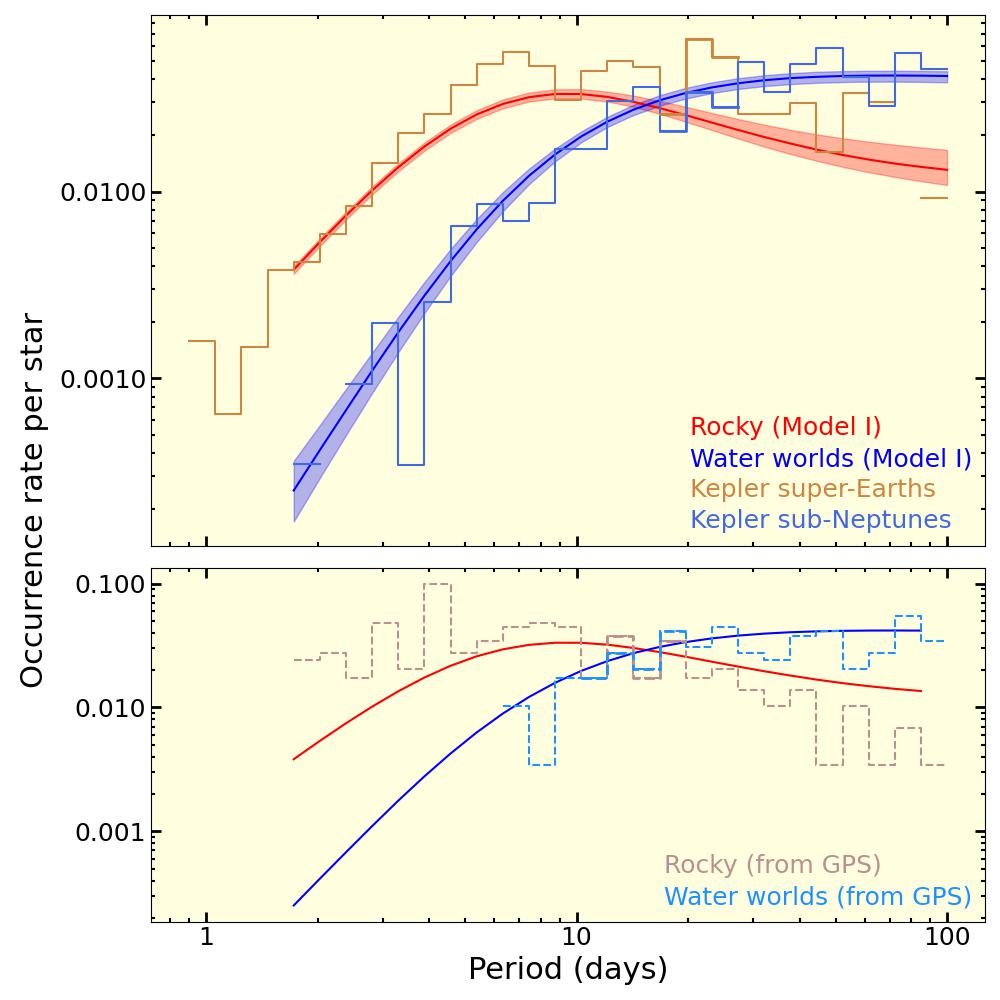}
\caption{Top: The period distributions of the rocky planets and water worlds from Model I compared to the completeness-corrected period distributions of the Kepler super-Earths and sub-Neptunes, respectively. The model period distributions are resampled from the MCMC posterior using Equations~\ref{eq:pkep}-\ref{eq:G-t}. The solid lines and the shaded regions (red and blue) denote the mean distributions and their 1$\sigma$ uncertainty ranges, respectively. Bottom: The mean period distributions from Model I are compared with those from a migration-based formation model from \textit{GPS} \citep{chakrabarty24, mulders20}, assuming a snowline location of 2.7 AU in the protoplanetary disk. Both the present model and the \textit{GPS} migration model suggest a consistent transition period.
}
\label{fig:pdist}
\end{figure}

We directly adopt the observed period distribution of Kepler planets as fixed input to our model and so, we do not compare the ensemble period distribution. Instead, we focus on comparing the period distributions of the individual rocky and water-rich populations with the Kepler super-Earths and sub-Neptunes, respectively. As shown in the top panel of Figure~\ref{fig:pdist}, Model I reproduces the observed distributions closely, even though the individual period distributions were not explicitly fit. The inferred parameters $\chi_1$ and $\chi_2$ are slightly higher and lower, respectively (deviating by more than $1\sigma$ but within $2\sigma$), compared to the values reported by \citet{bergsten22}, while $s$ is marginally higher but shows $1\sigma$ overlap. These differences can be attributed to our joint modeling of the period and radius distributions.

The distinct period distributions assumed for the rocky and water-rich planets align with predictions from planet formation and migration models. Rocky planets that form early and inside the snowline tend to migrate inward rapidly, populating the inner regions of the system \citep{mulders20, venturini20}. In contrast, the water-rich planets originating beyond the snowline often migrate more slowly and may stall at slightly longer orbital periods. This behavior may be influenced by dynamical processes such as trapping in mean-motion resonances with interior planets \citep{mulders20}, the outward movement of the magnetospheric cavity during disk dispersal \citep{liu17}, among others. 

To further assess the consistency, we use the migration-based population models from the \textit{Genesis} library\footnote{\url{https://github.com/GijsMulders/Genesis}} \citep{mulders20} for comparison. \textit{Genesis Population Synthesis}\footnote{\url{https://github.com/aritrachakrabarty/GPS}} \citep[GPS,][]{gps} identifies the rocky and water-rich populations from the ensemble population from the \textit{Genesis} library by calculating the WMF of those planets as a function of the snowline of the disk. The bottom panel of Figure~\ref{fig:pdist} compares the mean period distributions of the rocky and water worlds from Model I to those from \textit{GPS}. Although the distributions are not identical, the transition in the relative occurrence of rocky planets and water worlds with orbital period is broadly consistent between our current model (Model I) and \textit{GPS}. This transition is further found to be sensitive to the assumed location of the snowline in the protoplanetary disk, which in turn depends on the properties of the host star, such as the stellar mass \citep{mulders15} and stellar metallicity \citep{dominguez20}. This may potentially explain the increased occurrence of warm sub-Neptunes with increasing stellar metallicity \citep{mulders16, petigura18}. The bottom panel of Figure~\ref{fig:pdist} presents the case for a snowline at 2.7 AU for a G-type star, which produces the closest agreement with the transition period observed in the Kepler sample.

\begin{figure}
\centering
\includegraphics[scale=0.48]{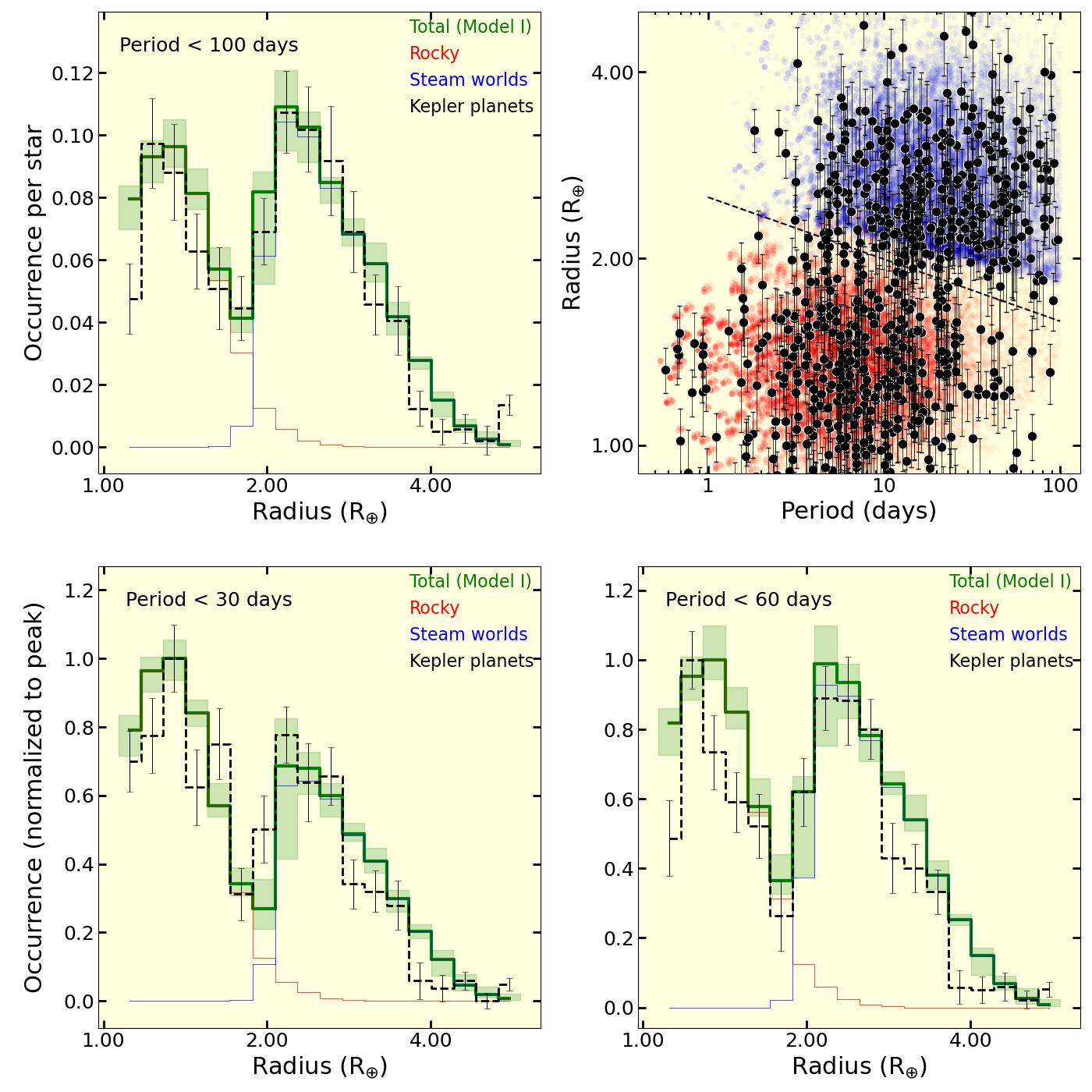}
\caption{Top left: Occurrence rates (per star) of the simulated rocky population, water world population, and total population from Model I, compared with the completeness-corrected occurrence rates of the Kepler planets from the CKS catalog in the complete analysis range of orbital periods ($P < 100$ days). The green solid line denotes the total occurrence rate corresponding to the median parameters, while the green shaded region indicates the 1-$\sigma$ uncertainty range. Top right: Period–radius distribution of the planets from Model I compared with that of Kepler planets from the CKS catalog. The model periods and radii are calculated from the MCMC posterior sample of the free parameters, with each point weighted by the Kepler detection completeness and its posterior likelihood. The dashed black line denotes the observed slope of radius valley over the orbital period \citep{vaneylen19}. Bottom panels: Same as the top left panel, but restricted to $P<30$ days and $P<60$ days, to better illustrate the period dependence of the radius distribution. The occurrence rates are normalized to their peaks, since we did not estimate $\ps{f}$, i.e., the average occurrence per star in these period ranges.
}
\label{fig:rp}
\end{figure}

\subsection{The radius histogram and period-radius distribution: primordial radius valley from dichotomy in bulk composition}\label{sec:res/rdist}

Figure~\ref{fig:rp} shows that our mixture model of rocky and water-rich populations from Model I efficiently reproduce both the radius histogram (top left panel) and the period–radius distribution (top right panel) of small Kepler planets. The top left panel displays the mean radius histogram and its 1$\sigma$ uncertainty band (in green), along with the individual contributions from the rocky planets (in red) and water worlds (in blue) to the total distribution. These are compared to the completeness-corrected radius histogram of Kepler planets. The top right panel presents the period–radius distributions of the model rocky and water-rich populations, computed from the MCMC posterior sample and weighted by their Kepler detection completeness and posterior likelihood. The separation between the two populations in the period–radius plane closely matches the estimated slope of the Kepler radius valley \citep{vaneylen19, gupta19}. Furthermore, to illustrate the joint distribution, the radius histograms for $P<30$ days and $P<60$ days are shown in the bottom panels, demonstrating good agreement with the observed planets in the same period ranges. The occurrence rates are peak-normalized, as \psm{f}~in these period ranges is not known.

Our results reinforce the notion that an \abin~compositional dichotomy can account for the radius valley, as proposed by previous studies \citep{izidoro22, burn24}. In particular, we demonstrate that a model based solely on steam worlds is capable of reproducing the entire observed radius histogram and period–radius distribution of Kepler sub-Neptunes, without invoking H/He envelopes and atmospheric loss models, for a reasonable range of model parameters. We notice a drop in the occurrence of such planets at 3-4 $\re$, since such large water worlds would require extremely high mass or WMF, suggesting the observed \textit{radius cliff} to be a result of this ``waterfall”. Planets larger than this are likely to possess significant H/He envelopes, indicating a distinct H/He-rich population that extends toward the regime of giant planets.

\subsection{The mass and WMF distributions}\label{sec:res/mwfdist}

\begin{figure}
\centering
\includegraphics[scale=0.5]{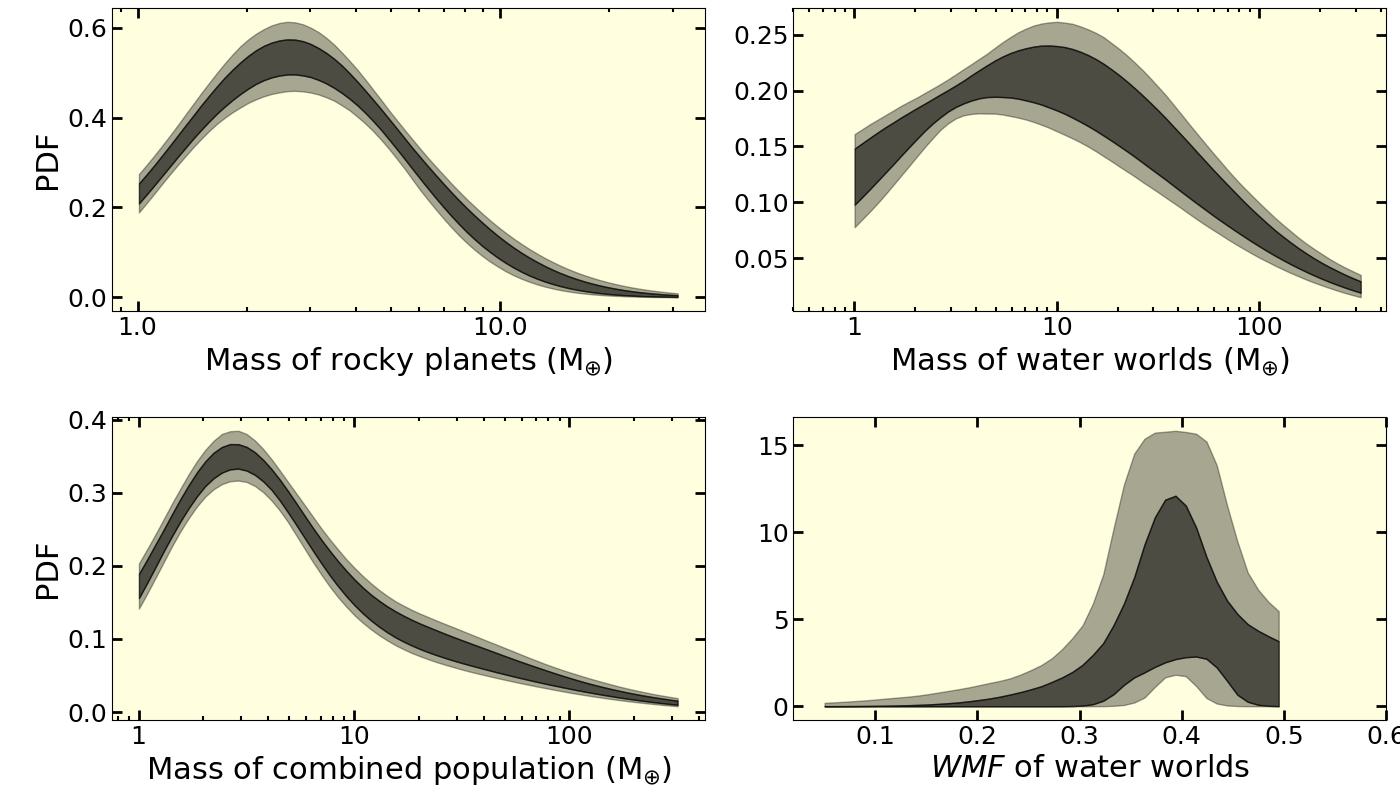}
\caption{Mass distributions of the rocky planets (top left), the water worlds (top right), and the total ensemble (bottom left), along with the WMF distribution of the water worlds (bottom right) from Model I. The dark and light bands represent the 1$\sigma$ and 2$\sigma$ uncertainty ranges, as computed from the MCMC posterior sample.
} 
\label{fig:mwf}
\end{figure}

We aim to infer the mass and WMF distributions of the small exoplanets from our model (Model I) that can reproduce the period-radius distribution of the Kepler short-period planets. The top panel of Figure~\ref{fig:mwf} shows the marginalized mass distributions for the rocky planets and water worlds. These distributions peak at approximately 2.6~$\me$ and 7~$\me$, respectively, indicating distinct populations. This separation is consistent with trends observed in the mass–radius distribution of exoplanets. We also investigate whether the combined mass distribution exhibits a bimodality---analogous to the radius distribution---that would further support the presence of two distinct populations. However, the bottom left panel of Figure~\ref{fig:mwf} shows that the total mass distribution is nearly unimodal, resembling that predicted by photoevaporation models of rocky cores with H/He envelopes \citep{wu19, rogers21}. This suggests that distinguishing between a planetary population shaped by a primordial compositional dichotomy and one sculpted by atmospheric evolution may remain challenging based on the observed mass and radius distributions.

The distribution of WMF, shown in the bottom right panel of Figure~\ref{fig:mwf}, peaks at $\sim$41\%, which is broadly consistent with the canonical 50\% value typically assumed for water worlds and with the water–ice fraction observed in the solar system \citep{lodders03}. The spread in the WMF distribution may reflect a range of factors, including variations in disk chemistry, differences in host star metallicity, or diverse accretion histories as planetary embryos migrate through the protoplanetary disk. It is also consistent with the model uncertainties in estimating the ice fractions in disks, e.g., the 38\% value for water ice among all condensates in the solar system from \citet{lodders03}, versus the 48\% estimated by \citet{anders89}.

\subsection{Mass-Radius relationship: A more realistic size limit of the steam worlds}\label{sec:res/mr}

\begin{figure}
\centering
\includegraphics[scale=0.55]{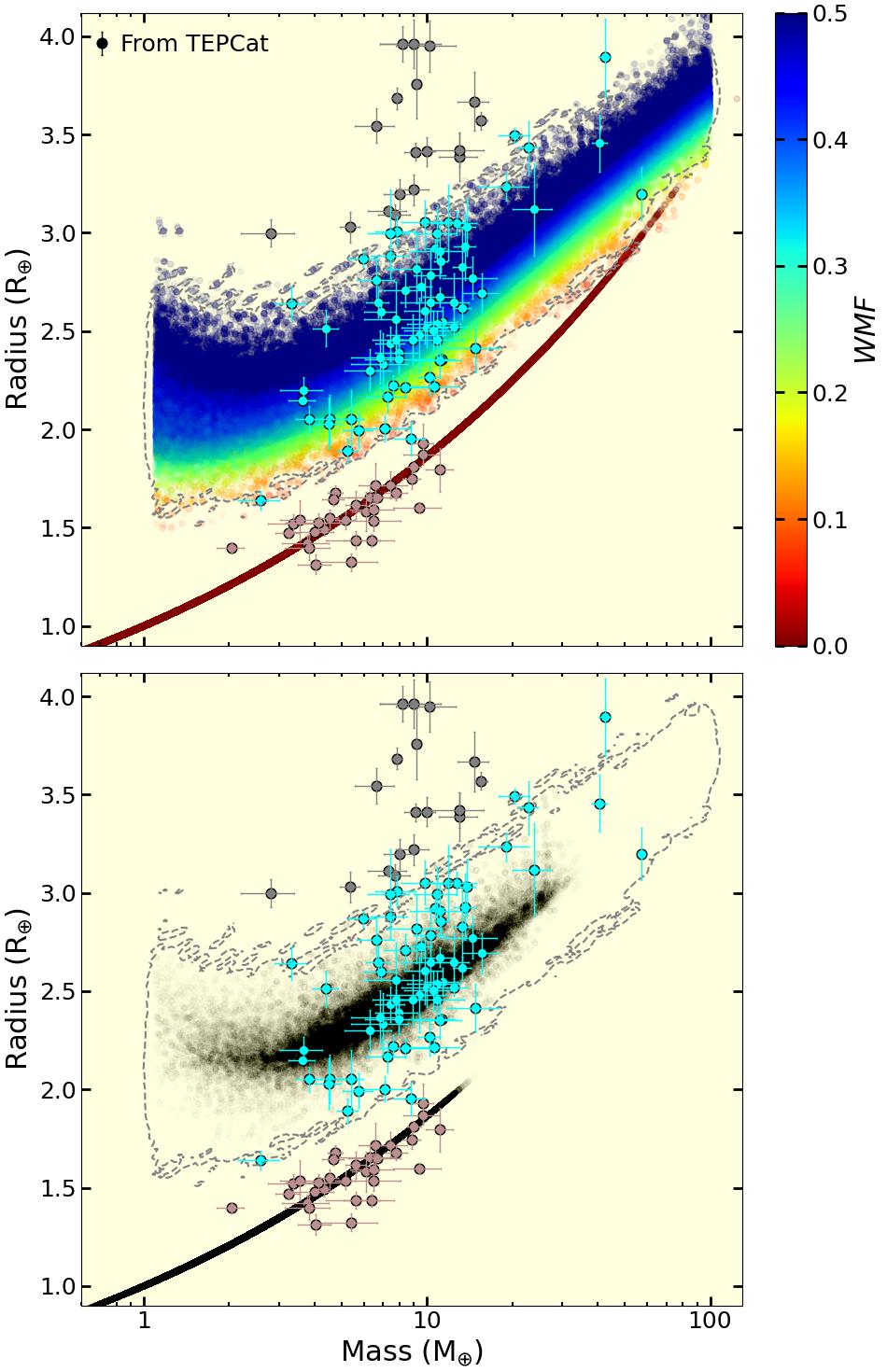}
\caption{Mass–radius distribution of planets from Model I compared with the observed planets around G-type host stars from the TEPCat catalog \cite{southworth11}. Top: Weighted mass–radius distribution of simulated rocky planets and water worlds, shown with a color map of water mass fraction (WMF). 
Bottom: The same weighted distribution shown without the WMF color map, with weight contours highlighted to illustrate regions of higher model likelihood and observational completeness. The model broadly agrees with the observed distribution for planets smaller than $\sim 3~\re$ in regions of high weights but fails to explain planets larger than $\sim 3~\re$, suggesting the presence of a H/He-rich population, comprising $\sim15\%$ of the full TEPCat sample and $\sim20\%$ of the TEPCat sub-Neptunes.}
\label{fig:mr}
\end{figure}

We compare our model (Model I) to the observed mass–radius distribution of small planets around G-type stars, defined by effective temperatures between 5200 K and 6500 K. We select planets from the TEPCat catalog with radii $\lesssim 4~\re$ and mass and radius measurements accurate to better than 25\% and 8\%, respectively, as shown with error bars in Figure~\ref{fig:mr}. Figure~\ref{fig:mr} also presents the mass–radius distributions of simulated rocky planets and water worlds, computed from the MCMC posterior and weighted by their posterior likelihood and Kepler detection completeness based on their radii and orbital periods. In the top panel, the model distribution is shown with a color map of WMF, but the effect of the weights is not visually apparent. Therefore, in the bottom panel we plot the same weighted mass–radius distribution without the color map, making the weight contours explicit.  

Although steam worlds can account for the radius distribution up to $\sim 4~\re$, the model underpredicts the observed low-density planets beyond $\sim 3~\re$, indicating a growing prevalence of gas dwarfs with substantial H/He envelopes at larger radii. We check whether the 1$\sigma$ error bars of the TEPCat planets fall within the water-world contour drawn using Model I (at 0.001 level, shown in grey dashes in Figure~\ref{fig:mr}), thereby classifying them as bare rocky, bare water-rich, and H/He-rich planets, shown in Figure~\ref{fig:mr} as brown, cyan, and grey error-bars, respectively. We estimate their overall occurrence fractions as 26.2\% rocky, 59\% water-rich, and 14.8\% H/He-rich, suggesting that at least $\sim20\%$ of the sub-Neptunes with both mass and radius measurements may be H/He-rich.

Our model predicts a significant population of low-mass sub-Neptunes that are underrepresented in the observed distribution due to the bias of RV surveys against detecting such planets. Conversely, although the outer grey contour of the water-world mass-radius distribution extends beyond $\sim50\me$, suggesting a finite simulated population of massive sub-Neptunes that are not observed, the probability density significantly drop in this regime (see also the bottom-left panel of Figure~\ref{fig:mwf}). In fact, such planets could instead be H/He-rich, in which case lower masses would be sufficient to explain their observed radii. Nevertheless, within regions of high density, the weighted model distribution broadly agrees with the observed population for planets with radii $\lesssim 3\re$, noting that this upper limit is actually mass dependent.

This highlights the importance of benchmarking models against the observed joint mass–radius–period distribution, underscoring the need for future surveys that provide comprehensive mass measurements with well-characterized detection completeness for small planets in the Kepler size range.

\section{Guidance to Future Observations}\label{sec:future}

A detailed demographic study across the full mass–radius–period parameter space can provide valuable constraints on planet formation and evolution models. Since a bias-corrected mass distribution for the Kepler population is not yet available, we use the computed total mass distribution of the simulated ensemble from Model I to help guide future observations aiming to measure the masses of small planets in the Kepler size range. We compare our model’s mass distribution with the completeness-corrected minimum mass distribution ($m \sin i$, where $i$ is the orbital inclination) from \citet{howard10}, who estimated planet occurrence rates for $m \sin i > 3~\me$ based on independent radial velocity (RV) observations. Although their mass histogram uses broad bins and reflects minimum rather than true masses, this comparison serves to contrast the mass patterns inferred from a model-based analysis of Kepler planets with those derived from independent RV surveys. We also compare our results with the bias-uncorrected mass distribution from the TEPCat catalog for G-type host stars (described in \S\ref{sec:res/mr}) to assess whether the model-predicted underlying mass distribution is reflected in the current sample of planets with precise mass and radius measurements. This planet sample is corrected only for Kepler detection completeness based on planet size and orbital period.

\begin{figure}
\centering
\includegraphics[scale=0.5]{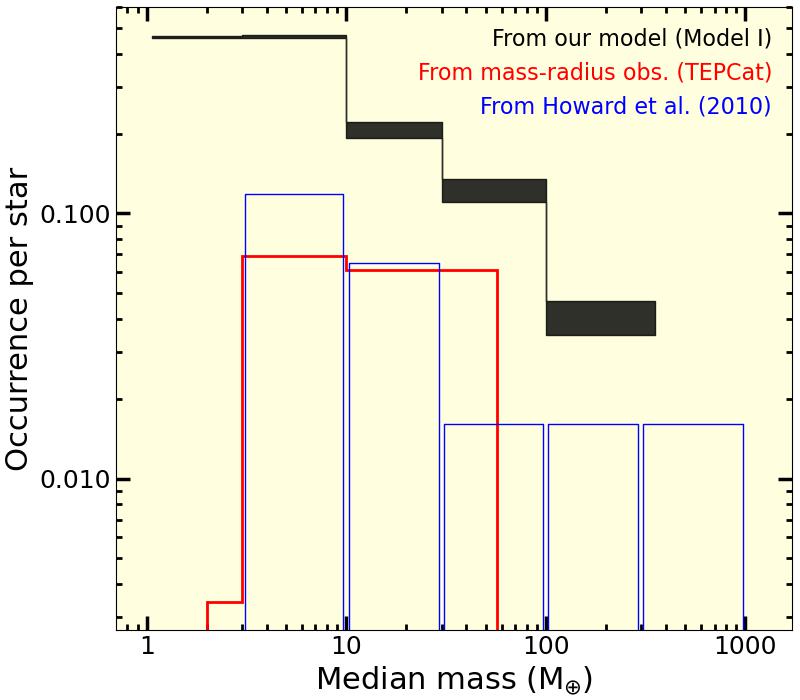}
\caption{
Comparison of the mass distribution from Model I with that from the TEPCat catalog (for G-type host stars) and the minimum mass ($m \sin i$) distribution from \citet{howard10}, all shown in the same mass bins as used by \citet{howard10}. The TEPCat distribution is corrected only for Kepler detection completeness based on planetary radius and orbital period. While Model I and the TEPCat distribution are restricted to small planets, the distribution from \citet{howard10} is not, since it is based on a pure RV sample with no measured radii. The high-mass tail from our model is found to be consistent with the observed distributions; because the biases preferentially enhance high-mass detections, this feature is unlikely to be spurious. However, the discrepancy in the occurrence rates between $\sim3$ and $\sim30 \me$ suggests that RV surveys likely underrepresent the Kepler population of small planets.
}
\label{fig:mdistobs}
\end{figure}

All three mass distributions—Model I, TEPCat, and \citet{howard10}—are shown in Figure~\ref{fig:mdistobs} using the same bins adopted by \citet{howard10}. While the overall shapes of the mass (or minimum mass) distributions are broadly consistent, the occurrence rates per star inferred from TEPCat and RV observations are significantly lower than those predicted by our model—by nearly 50\% in the 3–10~$\me$ range. This discrepancy suggests that RV surveys may underrepresent the Kepler population of 1–4 $R_\oplus$ planets.

A complementary avenue for identifying the origin of the radius valley is an age-based demographic study to determine whether this feature is primordial or the result of evolutionary processes. Recent studies using K2 and TESS data \citep{christiansen23, fernandes25} have attempted to estimate occurrence rates as a function of stellar age. These results indicate a decline in the overall occurrence rate of small exoplanets with increasing stellar age, suggesting the influence of evolutionary processes such as atmospheric loss. However, these estimates are limited by the small number of data points with large uncertainties in stellar ages, and planetary radii binned over broad intervals, which prevent clear distinctions between super-Earths and sub-Neptunes across age. Similarly, \citet{rogers25} conducts a simulation-driven investigation by generating synthetic TESS-biased young sub-Neptune populations under the ‘gas dwarf’ and ‘water world’ hypotheses and comparing them to observed TESS planets around stars younger than 40 Myr. His results tentatively disfavor the ‘water worlds’ hypothesis, though, as he points out, the small TESS sample limits the statistical significance. Therefore, the current observations cannot rule out a primordial origin driven by a bulk compositional dichotomy. Progress will require future surveys of larger, more uniform samples of young and intermediate-age stars, allowing a more precise characterization of age-dependent demographic studies of the super-Earths and sub-Neptunes. 

Finally, atmospheric spectroscopy provides an independent means of constraining the nature of sub-Neptunes on an individual-planet basis. Several sub-Neptunes with flat transmission spectra have nonetheless shown strong secondary-eclipse or phase-curve signals, suggesting high-metallicity, possibly steam-rich atmospheres. For instance, GJ 1214 b displays a featureless transmission spectrum, but JWST thermal-emission data reveal a steep slope consistent with a high mean molecular weight atmosphere, supporting a steam- or mixed-composition atmosphere over a purely H/He-rich envelope \citep{kempton23, gao23}. Conversely, a combined analysis of JWST and HST spectra of GJ 9827 d indicates absorption features of water and a metal-rich atmosphere, suggesting a steam atmosphere with a hydrosphere that may constitute up to $\sim 40\%$ of the planet’s mass \citep{piaulet24}.  Thus, coordinated transmission, emission, and phase-curve observations of sub-Neptunes can help disentangle clouds from intrinsically high-metallicity atmospheric compositions and potentially reveal whether these planets host steam- or mixed-composition atmospheres rather than purely H/He-rich envelopes.

Our analysis uses the static internal-structure model of \citet{aguichine21}, which can potentially overinflate the steam envelopes, especially for low-mass planets, as it does not account for thermal and structural evolution. Updated models that include evolutionary effects are now available \citep{aguichine25}, but incorporating them is beyond the scope of the present work and will be explored in future studies.

\section{Conclusion}\label{sec:con}

We model the Kepler planet population with radii between 0.9–6 \rem~and period (\p) $<$ 100 days as a mixture of rocky planets and water worlds using a Bayesian hierarchical framework. We find that migration-based formation scenarios are consistent with the observed transition from super-Earths to sub-Neptunes, likely reflecting a shift in dominant composition from rocky to water-rich planets across orbital periods. This offers an alternative explanation to the observed hot Neptune desert. The period-radius distribution of sub-Neptunes supports a water-rich composition, suggesting that the radius cliff is primarily shaped by the \textit{waterfall}---a sharp drop in the occurrence of water/steam-rich planets. At the same time, the observed mass–radius distribution from the TEPCat catalog shows an increasing prevalence of H/He-rich envelopes beyond $\sim 3 \re$. Based on a comparison between our model and planets with precise mass–radius measurements from the TEPCat catalog, we estimate that the small-planet population around G-type stars with orbital period $<100$ days comprises approximately 25\% rocky, 60\% water-rich, and 15\% H/He-rich planets. Our results also indicate that current RV surveys may underrepresent the planet population in terms of mass, and highlight the need for future missions with precise mass measurements and well-characterized completeness in the Kepler size range.

A.C. thanks Steve Bryson for insightful feedback on this work and acknowledges support from the NASA Postdoctoral Program at NASA Ames Research Center, administered by Oak Ridge Associated Universities under contract with NASA. G.D.M. acknowledges support from FONDECYT project 1252141 and the ANID BASAL project FB210003.

\bibliography{ref}

\appendix
\restartappendixnumbering 
\section{Corner plot for Model I}

\begin{figure}[h!]
\centering
\includegraphics[scale=0.3]{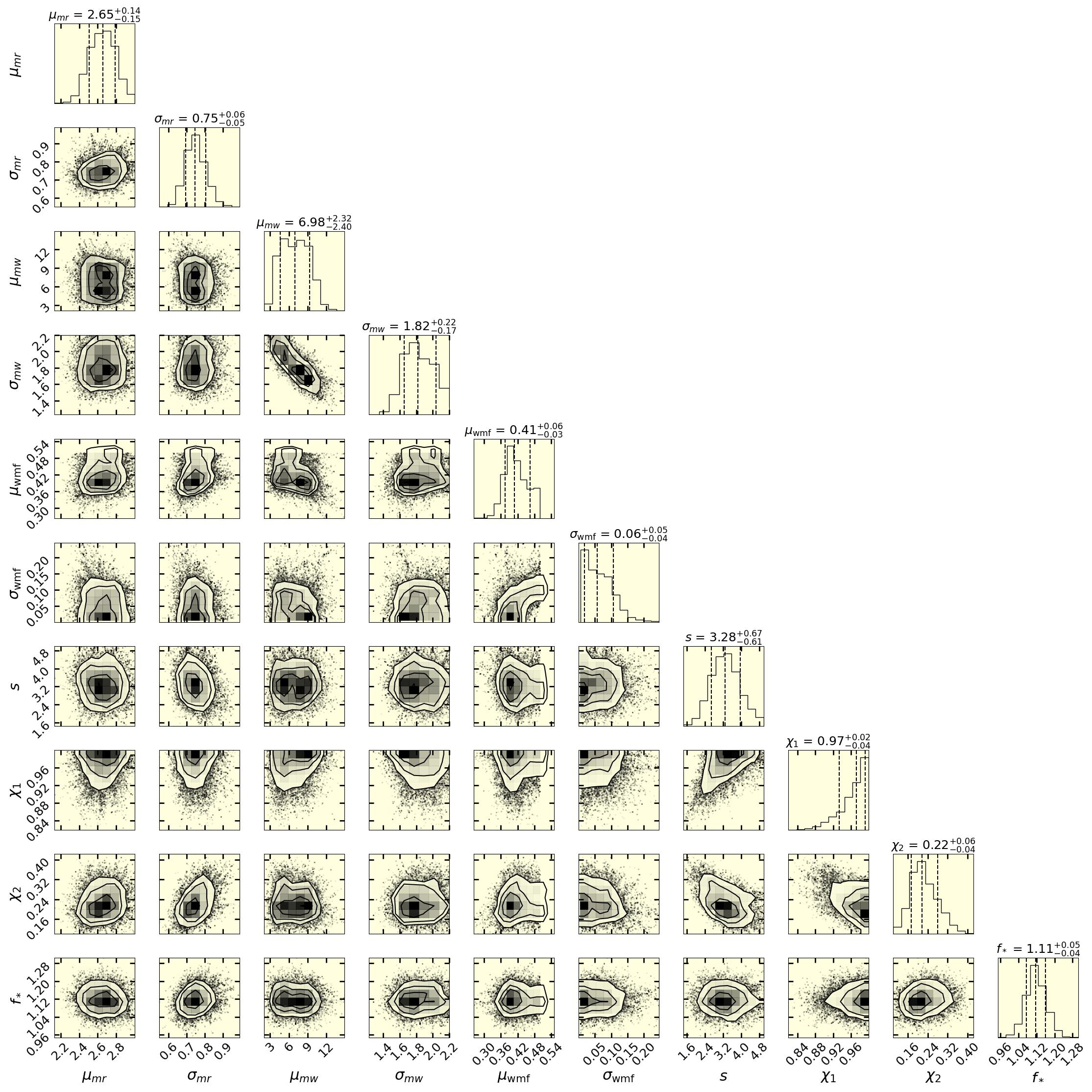}
\caption{
Corner plot showing the posterior distributions of the free parameters of Model I computed from the MCMC posterior sample. The quoted values represent the means and corresponding 1$\sigma$ uncertainties.
}
\label{fig:corner}
\end{figure}

\end{document}